\documentclass[a4paper,fleqn,usenatbib]{mnras}

\usepackage[T1]{fontenc}
\usepackage{ae,aecompl}

\usepackage{graphicx}	
\usepackage{amsmath}	
\usepackage{amssymb}	

\title[Zr, La, Ce and Eu]{Modelling the chemical evolution of Zr, La, Ce and Eu in the Galactic discs and bulge}

\author[Grisoni et al.]{V. Grisoni$^{1,2}$\thanks{E-mail: valeria.grisoni@inaf.it}, G. Cescutti$^{2,3}$, F. Matteucci$^{1, 2, 4}$, R. Forsberg$^5$, H. J{\"o}nsson$^{6,5}$,\newauthor N. Ryde$^{5,7}$\\
 $^1$ Dipartimento di Fisica, Sezione di Astronomia, Universit\`a di Trieste, via G.B. Tiepolo 11, I-34131, Trieste, Italy \\  
 $^2$ INAF Osservatorio Astronomico di Trieste, via G.B. Tiepolo 11, I-34131, Trieste, Italy\\
 $^3$ IFPU Institute for Fundamental Physics of the Universe, via Beirut 2, 34151, Trieste, Italy\\
 $^4$ INFN Sezione di Trieste, via Valerio 2, 34134 Trieste, Italy\\
 $^5$ Lund Observatory, Department of Astronomy and Theoretical Physics, Lund University, Box 43, SE-221 00 Lund\\
 $^6$ Materials Science and Applied Mathematics, Malm{\"o} University, SE-205 06, Malm{\"o}, Sweden\\
 $^7$ Universit\`e C$\hat{o}$te d'Azur, Observatoire de la C$\hat{o}$te d'Azur, CNRS, Laboratoire Lagrange, Bd de l'Observatoire, CS 34229,\\06304 Nice Cedex 4, France
}

\begin{document}
\date{Accepted . ; in original form xxxx}

\pagerange{\pageref{firstpage}--\pageref{lastpage}} \pubyear{xxxx}

\maketitle

\label{firstpage}

\begin{abstract}
We study the chemical evolution of Zr, La, Ce and Eu in the Milky Way discs and bulge by means of chemical evolution models compared with spectroscopic data. We consider detailed chemical evolution models for the Galactic thick disc, thin disc and bulge, which have been already tested to reproduce the observed [$\alpha$/Fe] vs. [Fe/H] diagrams and metallicity distribution functions for the three different components, and we apply them to follow the evolution of neutron capture elements. In the [Eu/Fe] vs. [Fe/H] diagram, we observe and predict three distinct sequences corresponding to the thick disc, thin disc and bulge, similarly to what happens for the $\alpha$-elements. We can nicely reproduce the three sequences by assuming different timescales of formation and star formation efficiencies for the three different components, with the thin disc forming on a longer timescale of formation with respect to the thick disc and bulge. On the other hand, in the [X/Fe] vs. [Fe/H] diagrams for Zr, La and Ce, the three populations are mixed and also from the model point of view there is an overlapping between the predictions for the different Galactic components, but the observed behaviour can be also reproduced by assuming different star formation histories in the three components. In conclusions, it is straightforward to see how different star formation histories can lead to different abundance patterns and also looking at the abundance patterns of neutron capture elements can help in constraining the history of formation and evolution of the major Galactic components.
\end{abstract}


\section{Introduction}

In the last years, several spectroscopic surveys have been developed and they have collected detailed stellar abundances of stars in the Milky Way, which are fundamental tools in order to reconstruct the history of formation and evolution of our Galaxy. Indeed, we are in a golden period for Galactic Archaeology, but still there are a lot of open questions that need to be answered by means of detailed theoretical models, such as for example the chemical evolution of neutron capture elements.
\\Neutron capture reactions were proposed by Burbidge et al. (1957) and Cameron (1957) to explain the origin of elements beyond Fe. In fact, chemical elements heavier than Fe cannot be the result of exoenergetic stellar fusion reactions. Instead, they must be synthesized by neutron capture on Fe-peak nuclei. The neutron capture process can be rapid (r-process) or slow (s-process) with respect to the $\beta$-decay timescale. Therefore, these elements are named r- and s-process elements, according to which of the two processes has contributed more to the production at solar metallicity.
\\For the s-process elements, the main production sites are suggested to be low-mass asymptotic giant branch (AGB) stars in the mass range 1.5-3.0 $M_{\odot}$ (Cristallo et al. 2009, 2011; Karakas 2010); these stars can build up all the neutron capture elements up to Pb and Bi and in this case the main source of neutrons is the reaction $^{13}$C($\alpha$,n)$^{16}$O. Also massive stars can provide neutron capture elements via s-process, but in this case the neutron flux is weaker and it comes from the reaction $^{22}$Ne($\alpha$,n)$^{25}$Mg; this is called "weak s-process", and generally the weaker neutron flux does not allow to produce very heavy elements, but only elements up to the magic number 50, for example Sr, Y and Zr.
\\For the r-process elements, an extremely neutron-rich environment is requested and in literature several production sites have been proposed. After the detection of the gravitational wave transient GW170817 (Abbott et al. 2017), neutron star mergers were greatly supported as production sites for r-process elements, but they might not be the only source (C{\^o}t{\'e} et al. 2019, Simonetti et al. 2019 and Bonetti et al. 2019). In literature, the first production sites proposed were core-collapse SNe (CC SNe) or electron-capture SNe (EC SNe) (Truran 1981; Cowan et al. 1991). However, Arcones et al. (2007) concluded that they do not have enough entropy and neutron fraction for an effective r-process activation. Thus, other production sites were suggested, such as neutron star mergers (NSM) (Rosswog et al. 1999) and magneto-rotationally driven SNe (MRD SNe) (Winteler et al. 2012; Nishimura et al. 2015).
\\From the point of view of Galactic chemical evolution (GCE) models, Matteucci et al. (2014) have explored the Eu production from NSM versus CC SNe. They concluded that NSM can account for the r-process enrichment in the Galactic halo, whether totally or partially in a mixed scenario with both Type II SNe and NSMs, giving a very short timescale of coalescence (but see Sch{\"o}nrich \& Weinberg 2019). Other studies have stated the importance of NSM in GCE models, but still NSM may not be the only source. Similar studies with GCE have investigated the scenario with MRD SNe (Cescutti \& Chiappini 2014) or the one with EC SNe (Cescutti et al. 2013). As regards to the s-process enrichment from GCE models, detailed studies were performed by Cescutti et al. (2006), Cescutti et al. (2013), Cescutti \& Chiappini (2014) and Cescutti et al. (2015), and they outlined the importance of s-process driven by rotation in massive stars. In these works, the nucleosynthesis prescriptions of Frischnecht et al. (2012, 2016) were used. Moreover, Prantzos et al. (2018) took into account the nucleosynthesis prescriptions of Limongi \& Chieffi (2018) (see Rizzuti et al. 2019 for a comparison between the yields of Frischnecht et al. 2016 and Limongi \& Chieffi 2018 in GCE models).
\\From the observational point of view, many studies have recently presented the abundance patterns of neutron capture elements in the different Galactic components, i.e. the Galactic halo, thick and thin discs, and bulge (Delgado-Mena et al. 2017, Forsberg et al. 2019). In particular, these studies show that in the [Eu/Fe] vs. [Fe/H] diagram, it is possible to see two distinct sequences, corresponding to the thick and thin discs stars (similarly to the [$\alpha$/Fe] vs. [Fe/H], see Hayden et al. 2015), at variance with other abundance patterns where the different populations are mixed (e.g. in the case of Zr, La and Ce). By studying the abundance patterns of different populations of stars at different metallicities, it is possible to understand which processes played a major role in the production of these elements at a given moment of the history of formation and evolution of our Galaxy, and Galactic chemical evolution models can shed light on that.
\\The aim of this paper is to study the chemical evolution of neutron capture elements (in particular, Zr, La, Ce and Eu) by means of detailed chemical evolution models in the light of the observational data from Forsberg et al. (2019). In particular, we consider the reference model of Grisoni et al. (2017) (see also Grisoni et al. 2018, 2019) for the Galactic thick and thin discs and the one of Matteucci et al. (2019) for the Galactic bulge; these models have been tested in order to reproduce the [$\alpha$/Fe] vs. [Fe/H] diagrams and metallicity distribution functions (MDF), and now we apply them to study the chemical evolution of neutron capture elements in order to shed light on the data by Forsberg et al. (2019).
\\This paper is structured in the following way. In Section 2, we present the observational data used in this work. In Section 3, we describe the chemical evolution models adopted. In Section 4, we discuss the results, based on the comparison between data and model predictions. Finally, in Section 5, we summarize the main conclusions.

\section{Observational data}

In this work, we use the data of Forsberg et al. (2019), where the chemical abundances of Zr, La, Ce, and Eu have been determined in 45 bulge giants and 291 local disc giants from high-resolution optical spectra.
\\The bulge spectra are obtained with the spectrometer FLAMES/UVES mounted on the VLT, Chile, with a resolution of R $\sim$ 47000. Five bulge fields are investigated, namely SW, B3, BW, B6, and BL after the naming scheme in Lecureur et al. (2007). The majority of the bulge stars are from the programs 71.B-0617, 73.B-0074 (PI: Renzini), observed in the years 2001-2003. This sample has been used in many works determining several abundances (Lecureur et al. 2007; Zoccali et al. 2006, 2008; Barbuy et al. 2013,2015; van der Swaelmen et al. 2016; da Silveira et al. 2018).
\\The bulk of the disc spectra in Forsberg et al. (2019) are obtained with the spectrometer FIES (Telting et al. 2014) at the Nordic Optical Telescope (NOT), La Palma (150 stars). Additional spectra were downloaded from the FIES archive (18 stars), added from Thygesen et al. (2012) (41 stars) and downloaded from the PolarBase data base (Petit et al. 2014) (19 stars). The PolarBase spectra are obtained with NARVAL and ESPaDOnS. The resolution of the disc spectra are R $\sim$ 67000 (FIES) and R $\sim$ 65000 (PolarBase). The disc sample has been separated into thick and thin disc components by using both chemistry and kinematics (Lomaeva et al. 2019).
\\The used wavelength region for abundance determination is restricted to that of the bulge spectra of 5800 \AA- 6800 \AA. The stellar parameters and abundances have been derived by fitting synthetic spectra using the code Spectroscopy Made Easy (SME, Valenti \& Piskunov 1996; Piskunov \& Valenti 2017). The stellar parameters for the same stellar sample is determined in J{\"o}nsson et al. (2017a,b). In da Silveira et al. (2018), the comparison of stellar parameters given between Zoccali et al. (2006) and Lecureur et al. (2007), and the parameters
derived in J{\"o}nsson et al. (2017a,b) is given.
\\The typical uncertainties on the determined abundances are around 0.08 dex for disc stars and 0.20 dex for bulge stars. For further details on the observational data and the determined abundances used in this work, we refer the reader to Forsberg et al. (2019), where there is also a detailed comparison with previous datasets present in the literature for the disc (Mishenina et al. 2013; Battistini \& Bensby 2016; Delgado-Mena et al. 2017; Guiglion et al. 2018) and the bulge (Johnson et al. 2012; van der Swaelmen et al. 2016; Duong et al. 2019).

\section{The models}

In this Section, we present the chemical evolution models used in this work. To follow the evolution of the Galactic thick and thin discs, we adopt the parallel approach (Chiappini 2009; Grisoni et al. 2017,2019; Cescutti \& Molaro 2019). In this approach, we consider that the thick and the thin disc formed by means of two distinct infall episodes and evolve separately. The model adopted here was developed for the solar neighborhood in Grisoni et al. (2017), and also tested for the other Galactocentric distances in Grisoni et al. (2018). For the Galactic bulge, we adopt the reference model of Matteucci et al. (2019), which considers a very short timescale of formation, higher star formation efficiency and flatter initial mass function (IMF) than the solar vicinity. These assumptions are required in order to reproduce the observed MDF of bulge stars, as first suggested by Matteucci \& Brocato (1990) and then confirmed also by subsequent theoretical studies (Ballero et al. 2007; Cescutti \& Matteucci 2011; Grieco et al. 2012; Cescutti et al. 2018; Matteucci et al. 2019). This corresponds to the so-called "classical bulge", but there can be other stellar populations coming via secular evolution from the inner disc (for a review on the chemodynamical evolution of the bulge, Barbuy, Chiappini \& Gerhard 2018 and references therein). The three Galactic components considered in this work thus differ by the different assumed IMF, timescales of gas accretion and efficiencies of star formation (see Table 1 for details). These assumptions have already been tested in previous works in order to reproduce the main observational features, such as the [$\alpha$/Fe] vs. [Fe/H] diagrams and MDF of disc stars (Grisoni et al. 2017, 2019) and bulge stars (Matteucci et al. 2019).

\subsection{Model equations}

The fundamental equations that describe the evolution with time of the mass fraction of the element $i$ in the gas $G_i$ are (see Matteucci 2012 for details):
\begin{align} \label{eq_01}
\begin{split}
& \dot G_i(r,t)= -\text{SFR}(r,t) X_i(r,t)+R_i(r,t)+ \dot G_i(r,t)_{inf}
\end{split}
\end{align}
with $\text{SFR}(r,t)$ being the star formation rate, $X_i(r,t)$ the abundance by mass of the element $i$, $R_i(r,t)$ the rate of matter restitution from stars of different masses into the interstellar medium (ISM) and finally $\dot G_i(r,t)_{inf}$ the rate of gas infall.
\\On the right hand side of Eq.\,\eqref{eq_01}, the first term corresponds to the rate at which the chemical elements are subtracted from the ISM for star formation processes. In this work, the SFR is expressed according to the Schmidt-Kennicutt law (Kennicutt 1998):
\begin{equation} \label{eq_03}
\text{SFR}(r,t)=\nu \sigma_{gas}^k(r,t),
\end{equation}
with $\sigma_{gas}$ being the surface gas density, $k=1.4$ the index of the law and $\nu$ the star formation efficiency, which is constrained in order to reproduce the SFR at the present time  in the considered Galactic component. In particular, for the Galactic bulge $\nu$ is very high compared to the ones of the thick and thin discs (see Table 1).
\\Then, in the term $R_i(r,t)$ of Eq.\,\eqref{eq_01}, we take into account detailed nucleosynthesis prescriptions from low and intermediate mass stars, Type Ia SNe (originating from white dwarfs in binary systems) and Type Ib, Ic and II SNe (originating from core-collapse massive stars). In this work, we consider also the contribution from NSM, which are fundamental europium producers. In the term $R_i(r,t)$, the IMF is involved (see Matteucci 2012 for details). Here, we adopt the Scalo (1986) IMF for the Galactic thick and thin discs, and the Salpeter (1955) IMF for the Galactic bulge. In fact, the IMF for the bulge should be flatter than the one adopted for the solar neighbourhood, see Matteucci et al. (2019).
\\The last term in Eq.\,\eqref{eq_01} refers to the rate of gas infall. In particular:
\begin{equation} \label{eq_02}
\dot G_i(r,t)_{inf}=A(r)(X_i)_{inf}e^{-\frac{t}{\tau}},
\end{equation}
with $G_i(r,t)_{inf}$ being the infalling material in the form of the element $i$ and $(X_i)_{inf}$ the composition of the infalling gas, which is assumed to be primordial. The parameter $\tau$ represents the timescale of mass accretion in each Galactic component; the timescales are free parameters of the models and they are constrained mainly by the comparison with the observed MDF. The quantity $A(r)$ is a parameter obtained by reproducing the total surface mass density at the present time in the considered Galactic component. We follow the prescriptions of Grisoni et al. (2017) for the Galactic discs and Matteucci et al. (2019) for the Galactic bulge.

\begin{table*}
\caption{Input parameters for the chemical evolution models used in this work. In the first column, we indicate the name of the model. In the second column, there is the assumed initial mass function. In the third column, we indicate the star formation efficiency ($\nu$). In the fourth column, we give the timescale for mass accretion ($\tau$).}
\label{tab_01}
\begin{center}
\begin{tabular}{c|cccccccccc}
  \hline
 Model & IMF &$\nu$& $\tau$\\
&  &[Gyr$^{-1}$]& [Gyr]\\
 \hline

Thin disc & Scalo (1986) & 1.2  & 7 \\

 \hline

Thick disc & Scalo (1986) & 2& 0.1\\

 \hline

Bulge & Salpeter (1955) & 20  &0.1 \\

 \hline

\end{tabular}
\end{center}
\end{table*}

\begin{figure*}
\includegraphics[scale=0.45]{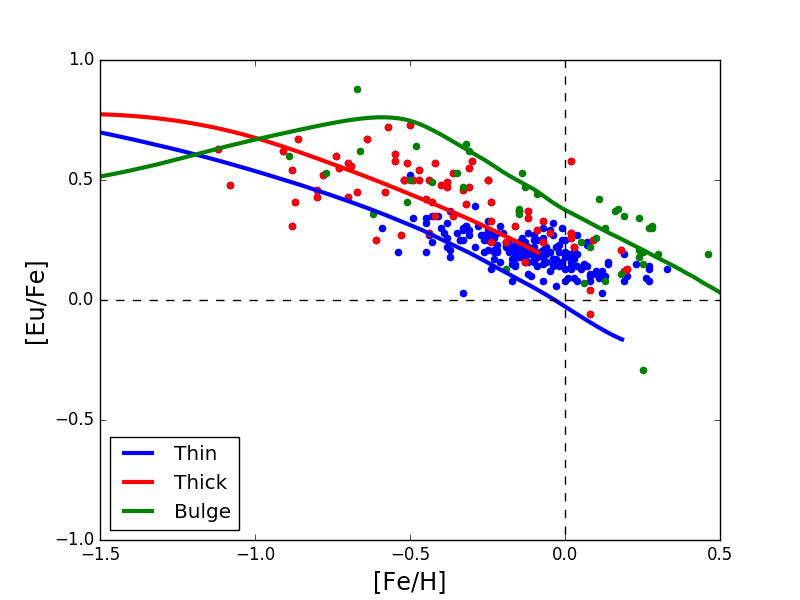}
 \caption{Observed and predicted [Eu/Fe] vs. [Fe/H]. The predictions are from the reference models for the Galactic thin disc (blue line), thick disc (red line) and bulge (green line). The data are for the Galactic thin disc stars (blue dots), thick disc stars (red dots) and bulge stars (green dots), and they are taken from Forsberg et al. (2019).}
 \label{fig_01}
\end{figure*}

\begin{figure}
\includegraphics[scale=0.45]{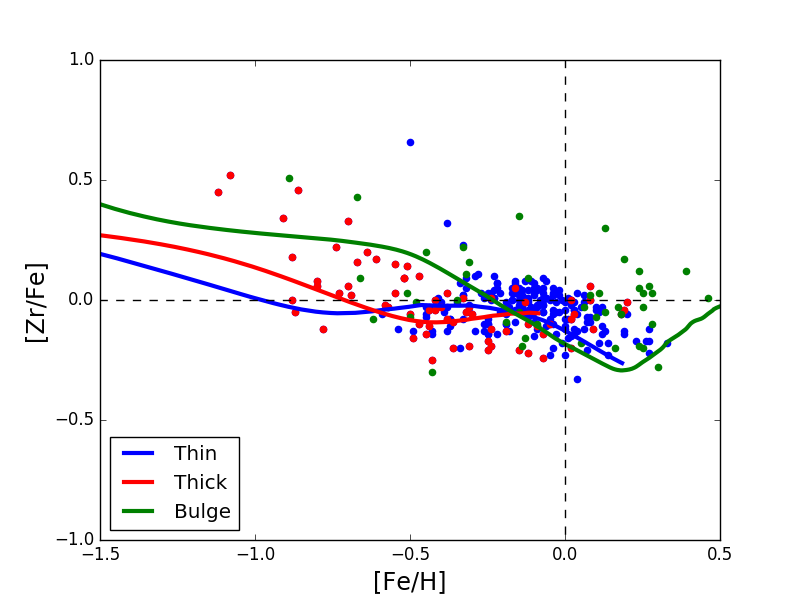}
\includegraphics[scale=0.45]{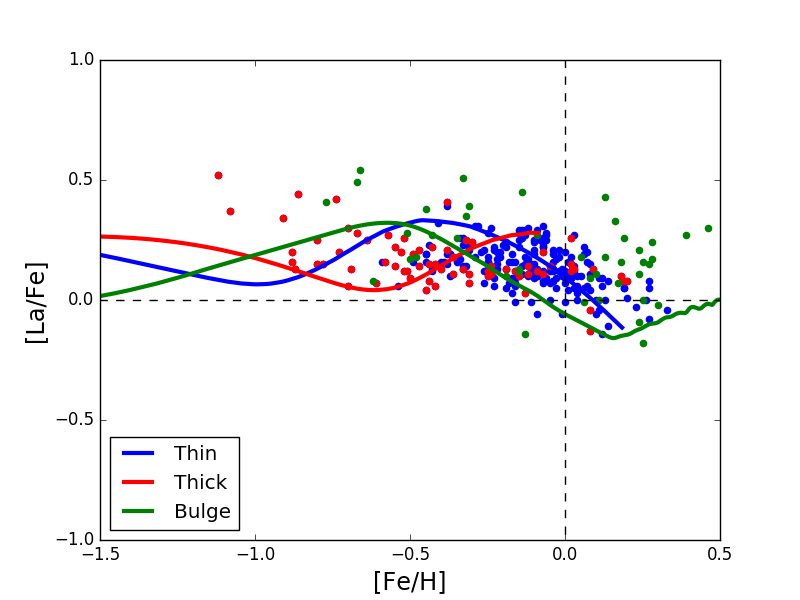}
\includegraphics[scale=0.45]{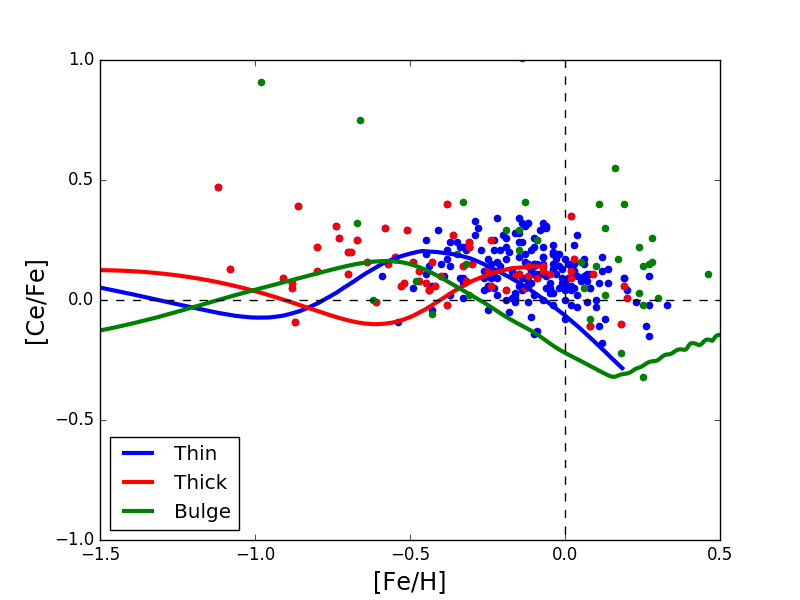}
 \caption{Same as Fig. 1, but for Zr, La and Ce.}
 \label{fig_02}
\end{figure}

\subsection{Nucleosynthesis prescriptions}

In this work, we adopt the following nucleosynthesis prescriptions for Zr, La, Ce and Eu.

\subsubsection{Yields of Zr, La and Ce}

Zr, La and Ce are produced by both the r- and s- processes.
\\The r-process yields are obtained by scaling the Eu yields according to the abundance ratios observed in r-process rich stars (Sneden et al. 2008).
\\Low-mass AGB stars in the mass range 1.3-3 M$_{\odot}$ are responsible for most of the s-process, and the corresponding yields are taken from the database FRUITY (FUll-Network Repository of Updated Isotopic Tables \& Yields, Cristallo et al. 2009, 2011).
\\Then, we assume also the s-process contribution from rotating massive stars. This has been first considered by Cescutti et al. (2013), Cescutti \& Chiappini (2014) and Cescutti et al. (2015) by taking into account the nucleosynthesis prescriptions of Frischknecht et al. (2012). Here, we consider the nucleosynthesis prescriptions of Frischknecht et al. (2016) for rotating massive stars.

\subsubsection{Yields of Eu}

For Eu, we consider NSM as fundamental production sites, as mentioned in the Introduction. To include the production of Eu from NSM in the Galactic chemical evolution models, we need to define the following quantities (see Matteucci et al. 2014, Cescutti et al. 2015 for further details):
\begin{itemize}
\item the fraction of massive stars belonging to double neutron star systems that will eventually merge, or in other words the realization probability of such events ($\alpha_{NSM}$);
\item the time delay between the formation of the double neutron star system and the merging ($\Delta$t$_{NSM}$);
\item the amount of Eu produced during the merging (M$^{Eu}_{NSM}$).
\end{itemize}
Concerning NSM yields, we follow the prescriptions of Matteucci et al. (2014) and Cescutti et al. (2015); in particular, we assume a value of 2$\cdot$10$^{-6}$ M$_{\odot}$ which is in agreement with the range of yields of Korobkin et al. (2012) who suggest that NSM can produce from 10$^{-7}$ to 10$^{-5}$ M$_{\odot}$ of Eu per event.
\\We assume that a fixed fraction of all the massive stars is a progenitor of NSM and produces r-process material. The progenitors are randomly chosen among the massive stars formed in the stellar mass range 10-30 M$_{\odot}$. The parameter $\alpha_{NSM}$ is taken equal to 0.05, in order to reproduce the present time rate of NSM in the Galaxy as given by Kalogera et al. (2004) (R$_{NSM}$=83$^{+209}_{-66}$Myr$^{-1}$). The recent observations of the rate for the event GW170817 seem to confirm this result (Matteucci et al. 2018).
\\For the time delay due to the coalescence of the two neutron stars, it is fixed and equal to 1 Myr as in Matteucci et al. (2014) and Cescutti et al. (2015) (which is very short, but see Sch{\"o}nrich \& Weinberg 2019 which have allowed for a 2-phase ISM in order to solve this problem). It is worth noting also that here it is assumed that all neutron star binaries have the same coalescence time, but a more realistic approach would consider a distribution function of such timescales, in analogy with SNIa for which a distribution for the explosion time is defined (see Simonetti et al. 2019). The aforementioned assumptions on NSM are probably extreme concerning the short and constant merging timescale, but they reproduce very well the [Eu/Fe] vs. [Fe/H] plot in the solar neighborhood. On the other hand, both Matteucci et al. (2014) and Cescutti et al. (2015) have demonstrated that they can obtain a good agreement with the data also by assuming CC SNe producing Eu at early times and larger merging timescales.
\\In the model, Eu is produced also by the main s-process, but this is only the 5$\%$ fraction of the total abundance, and therefore NSM remain the main source of Eu.

\subsubsection{Yields of Fe}

Finally, the iron yields are the ones of Kobayashi et al. (2006) for CC SNe and Iwamoto et al. (1999) for SNIa.

\section{Comparison between data and model predictions}

In this Section, we show our results based on the comparison between model predictions and observations for the various Galactic components: thick disc, thin disc and bulge. In Table 1, the input parameters of the different models are listed. In the first column, there is the name of the model. Then, we indicate the adopted IMF, the star formation efficiency ($\nu$) and the timescale of formation ($\tau$) of the Galactic components.

\subsection{Comparison for Eu}

In Fig. \ref{fig_01}, the observed and predicted [Eu/Fe] vs. [Fe/H] plot is shown in the range -1.5 < [Fe/H] < 0.5 dex. The trend shows a plateau at [Fe/H] < -0.6 dex and then a decrease with metallicity, and we can see that there are three distinct sequences in this diagram, corresponding to the three main Galactic components: thick disc, thin disc and bulge. The bulge is Eu-enhanced with respect to the disc (thick + thin). The bulge abundances are indeed higher than the thin disc, but it is fairly hard to tell if the bulge is higher in abundance than the thick disc due to the larger scatter in the bulge data (Forsberg et al. 2019). As regards to the disc, it clearly shows a dichotomy between the thick and thin disc stars, with the thick disc been Eu-enhanced with respect to the thin disc. The models can nicely reproduce the observations, by assuming different timescales of formation and star formation efficiencies in the three Galactic components. In particular, the bulge has formed on a short timescale of formation and with high star formation efficiency, and the thick disc has a shorter timescale of formation and higher star formation efficiency than the thin disc (see Table 1). The adopted input parameters were previously tuned in order to reproduce the observed [$\alpha$/Fe] vs. [Fe/H] and MDFs for the Galactic thick and thin discs (Grisoni et al. 2017) and for the Galactic bulge (Matteucci et al. 2019). Now, we see that these parameters can nicely fit also the [Eu/Fe] vs. [Fe/H] relation.
\\The model for the Galactic thin disc correctly reproduces the solar value, as expected. Regarding the solar value, the observational data seem to be overestimated. In Forsberg et al. (2019), it is indeed noted that the Eu (as well as the La and the Ce) abundances might suffer from systematic errors, causing an overestimation in abundances. Their differential comparison of the disc and bulge components are therefore not affected directly by the overestimation, but should be considered when comparing to our models.
\\Moreover, we note that the observed [Eu/Fe] vs. [Fe/H] in the thin disc flattens at high metallicities (see also Delgado-Mena et al. 2017). This could be due to radial migration (Sch{\"o}nrich \& Binney 2009; Minchev et al. 2013, 2018; Spitoni et al. 2015) from the inner disc, similarly to what happens in the [Mg/Fe] vs. [Fe/H] plot (Grisoni et al. 2017). Recently, the importance of radial migration in shaping also the r-process abundance pattern has been investigated by Tsujimoto \& Baba (2019).
\\Concerning the Galactic thick disc, it is characterized by a more intense star formation history than the thin disc. There is a faster evolution, with a stronger efficiency of star formation ($\nu$=2 Gyr$^{-1}$) and shorter timescale of gas infall ($\tau$=0.1 Gyr). The chemical evolution of the thick disc lasts for 2 Gyr, but with minimal star formation after approximately 1.2 Gyr. Thus, the model predictions for the thick discs stop at [Fe/H]$\sim$ -0.1 dex. Overall, the predictions for the thick disc lie above the ones for the thin disc, and this is due to the much faster evolution.
\\A first attempt to reproduce the distinct sequences in the [Eu/Fe] vs. [Fe/H] plot has been performed in Delgado-Mena et al. (2017) by means of the chemical evolution models of Bisterzo et al. (2017). In that case, the data for the thin disc were nicely reproduced, but the predictions for the thick disc were underestimated with respect to the ones of the thin disc, at variance with the observations. In order to correctly reproduce the observed dichotomy between the thick and thin disc stars, we need to assume that the thick disc formed with a shorter timescale of formation and higher star formation efficiency than the thin disc. Moreover, here we present the predictions of the [Eu/Fe] vs. [Fe/H] in the Galactic bulge.
\\Concerning the Galactic bulge, here we consider the model of Matteucci et al. (2019), which assumes an even faster and more efficient evolution ($\nu$=20 Gyr$^{-1}$, $\tau$=0.1 Gyr) and a flatter IMF (Salpeter 1955) with respect to the solar vicinity. This model reproduces the [$\alpha$/Fe] vs. [Fe/H] diagram and MDF of bulge stars, and it corresponds to the "classical bulge". However, we remind that in the Galactic bulge there is the possibility of other stellar populations originating via secular evolution from the inner disc and coexisting with the bulge stars formed in situ, but firm conclusions are still not reached at this point. Here, we consider only the classical bulge population, which should be the dominant one. Due to its faster and more efficient formation, the track for the Galactic bulge is Eu-enhanced with respect to the thick and thin discs. However, we notice that the predicted [Eu/Fe] at [Fe/H]=-1.5 dex starts from a lower value than the other Galactic components, but then the track rises and then it decreases with metallicity being Eu-enhanced with respect to the other Galactic components. The knee is thus shifted towards higher metallicities with respect to the thick and thin discs, and this can be explained in terms of the so-called time-delay model (Matteucci 2012). In fact, the bulge forms on a shorter timescale of formation and with a higher star formation efficiency than the other Galactic components, and therefore its knee is shifted towards higher metallicities.
\\In summary, by assuming different star formation histories which have already allowed us to reproduce the abundance patterns of the $\alpha$-elements in the Galactic thick and thin discs (Grisoni et al. 2017) and bulge (Matteucci et al. 2019), we can nicely reproduce the three sequences also in the [Eu/Fe] vs. [Fe/H] plot.

\subsection{Comparison for Zr, La and Ce}

Now, we present the results for the other chemical elements of this study, i.e. Zr, La and Ce.
\\In Fig. \ref{fig_02}, we show the observed and predicted [X/Fe] vs. [Fe/H] for Zr, La and Ce. As explained in Section 3.2, these chemical elements are produced by both the s- and r- processes, and they show a different abundance pattern than Eu. In particular, the predominant s-process fraction of these elements is produced by long-lived stars (1.5-3.0 $M_{\odot}$). This fact, coupled with the secondary nature (dependence on metallicity) of s-process elements creates the behaviour observed in Fig. \ref{fig_02}. Moreover, Zr is a first-peak s-process element, whereas La and Ce are second-peak s-process elements, and indeed the predictions for Zr are slightly different than the ones for La and Ce, which show a more similar behaviour. In these diagrams, the three stellar populations (thin disc, thick disc and bulge) are mixed, and it is more difficult to disentangle the different behaviours. However, we can see that the general behaviour of the observational data for each Galactic component is reproduced by the models, with the thin disc showing a decrease with increasing [Fe/H], at variance with the bulge that shows a slight increase at higher metallicities, whereas the thick disc represents an intermediate case. The different behaviours of the s-process elements in the bulge is due to the time-delay model. In fact, it has a regime of high star formation rate and thus the curve for the thick disc should be shifted towards the right in the [s/Fe] vs. [Fe/H] plot. Also in these diagrams, assuming different timescales of formation and star formation efficiencies can lead to different behaviours in the abundance patterns, even if the different populations are mixed and it is more difficult to disentangle the different patterns at variance with the [Eu/Fe] vs. [Fe/H] plot where three distinct sequences are evident.
\\In conclusion, also looking at the abundance patterns of neutron capture elements in the Milky Way discs and bulge can help in constraining the history of formation and evolution of these three Galactic components.

\section{Conclusions}

In this paper, we have studied the chemical evolution of Zr, La, Ce and Eu in the Galactic discs and bulge by means of detailed Galactic chemical evolution models compared with the data by Forsberg et al. (2019).
\\The main conclusions of this work can be summarized as follows.
\begin{itemize}
\item In the [Eu/Fe] vs. [Fe/H] plot, we observe and predict three distinct sequences, corresponding to the Galactic thick disc, thin disc and bulge (similarly to what happens in the [$\alpha$/Fe] vs. [Fe/H] plot, see Grisoni et al. 2017, Matteucci et al. 2019).
\item The three sequences in the [Eu/Fe] vs. [Fe/H] plot are reproduced by assuming three different star formation histories for these Galactic components, with the bulge forming on a shorter timescale of formation and with higher star formation efficiency than the discs. Moreover, the thick and thin discs show a clear dichotomy, with the thick disc forming faster than the thin one, in agreement with the results of Grisoni et al. (2017).
\item The assumed timescales of gas infall and star formation efficiencies have been previously tuned to reproduce the [$\alpha$/Fe] vs. [Fe/H] plots and the MDFs of the thick and thin discs (Grisoni et al. 2017) and bulge (Matteucci et al. 2019), and they allow us to nicely reproduce also the [Eu/Fe] vs. [Fe/H] plot.
\item On the other hand, we observe and predict a different behaviour for the [X/Fe] vs. [Fe/H] plots of Zr, La and Ce. This is due to the double nature of these elements, which are produced by either the s- and r- processes. In fact, Zr, La and Ce are mainly produced as s-process elements by low mass stars (1.5-3.0 $M_{\odot}$) and only partly as r-process. As it is well known, the s-process elements behave as secondary elements. This fact, coupled with the long timescale of their production, produce the increase with metallicity followed by a decline.
\item In the [X/Fe] vs. [Fe/H] plots for Zr, La and Ce, the three stellar populations are mixed and it is more difficult to disentangle them. However, the general behaviour of the observational data also in this case can be reproduced by the models and interpreted in terms of the time-delay model.
\end{itemize}
In conclusion, in addition to the study of the abundance patterns of $\alpha$-elements in the Galactic discs and bulge, also looking at the abundance patterns of neutron capture elements can help in constraining the history of formation and evolution of these three Galactic components.

\section*{Acknowledgments}

V.G. and F.M. acknowledge financial support from the University of Trieste (FRA2016).
\\G.C. acknowledges financial support from the European Union Horizon 2020 research and innovation programme 
under the Marie Sk\l{}odowska-Curie grant agreement No. 664931 and from the EU COST Action CA16117 (ChETEC).
\\H.J. acknowledges support from the Crafoord Foundation, Stiftelsen Olle Engkvist Byggm\"astare, and Ruth och Nils-Erik Stenb\"acks stiftelse.
\\N.R. acknowledge support from the Swedish Research Council, VR (project numbers 621-2014-5640), and the Royal Physiographic Society in Lund through Stiftelsen Walter Gyllenbergs fond and M{\"a}rta och Erik Holmbergs donation.
\\We thank the referee for the insightful comments and suggestions.

\end{document}